\documentclass[aps,prl,reprint,groupedaddress]{revtex4-2}

\usepackage{graphicx}
\usepackage{tikz}
\usepackage{epstopdf}
\usepackage{subfigure}
\usepackage[percent]{overpic}

\usepackage{amsmath,amssymb,amsfonts}%
\usepackage{amssymb}   
\usepackage{amsthm}%
\usepackage{mathrsfs}%

\usepackage{lineno}
\usepackage{hyperref}
\hypersetup{
	colorlinks,
	citecolor=black,
	filecolor=black,
	urlcolor=black
}

\begin{document}


\title{First Observation of Fishbone-Driven Zonal Flows with Fine Reversed Structure \\ in Tokamak Plasmas}


\author{Liutian Gao,$^{1,*}$
Yuehao Ma,$^{1,*}$
Huishan Cai,$^{1,\dagger}$
Adi Liu,$^{1,\dagger}$
Bin Zhang,$^{2}$ 
Ming Xu,$^{2}$ Haiqing Liu,$^{2}$ Liqing Xu,$^{2}$ \\Chu Zhou,$^{1}$ Feifei Long,$^{1}$  
Mingyuan Wang,$^{1}$ Xiaoming Zhong,$^{2}$ Jinlin Xie,$^{1}$ Ge Zhuang,$^{1}$ and the EAST team\\
}
\affiliation{\vspace{3pt}\\
$^1$School of Nuclear Science and Technology, University of Science and Technology of China, Hefei 230026, China\\
$^2$ Institute of Plasma Physics, Chinese Academy of Sciences, Hefei 230031, China\\
\textup{${}^*$ These authors contributed equally to this work\\
${}^\dagger$ Corresponding author email: hscai@mail.ustc.edu.cn, lad@ustc.edu.cn}
}


\date{\today}

\begin{abstract}
We present the first direct experimental observation of 
fishbone-driven zonal flows 
in the core of the EAST tokamak. 
In contrast to the global pattern predicted 
by previous models and simulations based on the energetic-particle-expulsion mechanism, 
the observed flows exhibit a fine-scale, radially reversed structure 
inside the $q = 1$ rational surface. 
The flow rises faster and saturates earlier than the fishbone within a single burst, 
indicating that a beat-driven nonlinear process dominates the early stage 
rather than the energetic-particle-expulsion mechanism. 
Global nonlinear gyrokinetic simulations quantitatively reproduce the observed radial profile 
and reveal that this structure arises from the cancellation of comparable 
but opposite contributions from thermal ions and electrons.  
This cancellation mechanism is not captured in previous theoretical frameworks. 
These findings establish that the fishbone can generate sheared flows with a distinct radial topology, 
offering a promising pathway for regulating turbulence and improving core confinement.

\end{abstract}


\maketitle

\paragraph{\label{sec:introduction}Introduction.---}

Sheared flows are ubiquitous in nature, 
appearing in planetary atmospheres 
\cite{porcoCassiniImagingJupiters2003}, 
astrophysical accretion disks 
\cite{balbusInstabilityTurbulenceEnhanced1998}, 
and magnetically confined fusion plasmas
\cite{Terry2000,diamondZonalFlowsPlasma2005}. 
In tokamaks, sheared $\boldsymbol{E} \times \boldsymbol{B}$ flows, 
particularly zonal flows (ZFs), 
play a crucial role in regulating plasma instabilities 
\cite{Biglari1990,Diamond1991,burrellEffectsExBVelocity1997,Lin1998,Connor2004,fujisawaReviewZonalFlow2009,Ida2018,yoshida2025transport}, 
such as turbulence and modes driven by energetic particles (EPs). 
They suppress turbulent transport and thereby improve confinement 
\cite{diamondSelfRegulatingShearFlow1994,xuRoleReynoldsStressInduced2000,kimZonalFlowsTransient2003,Schmitz2012}, 
while also modulating the nonlinear saturation of EP-driven instabilities 
and the resulting EP transport \cite{qiuEffectsEnergeticParticles2016,qiuFineStructureZonal2016a,qiuNonlinearExcitationFiniteradialscale2017,chenZonalStructureEffect2018,Brochard2024}. 
In burning plasmas, microturbulence and EP-driven instabilities, 
including alpha-particle-driven modes, can interact strongly. 
ZFs generated by EP-driven instabilities have been identified as 
a key mediator of this cross-scale interaction 
\cite{liuRegulationAlfvenEigenmodes2022,liuCrossscaleInteractionMicroturbulence2024,Ma2026crossscale}. 
Therefore, a systematic investigation of the generation and structure 
of ZFs induced by EP-driven instabilities is essential for understanding confinement and transport in burning plasmas.

The fishbone instability is a common macroscopic 
EP-driven instability in tokamak plasmas
and is considered an important issue for ITER-relevant
baseline and hybrid scenarios \cite{Mcguire1983study,Chen1984excitation,Coppi1986theoretical,salewskiEnergeticParticlePhysics2025}. 
Fishbone bursts have been experimentally observed to correlate with the formation of internal transport barriers (ITBs) in several tokamaks, 
including ASDEX Upgrade \cite{Gunter2001mhd}, HL-2A \cite{Chen2016dynamics,Deng2022investigation}, 
EAST \cite{Gao2018sustained,Liu2020experimental,Zhang2022} and DIII-D \cite{brochard2025}.
A long-standing hypothesis is that fishbone-driven flows, generated through EP expulsion or redistribution, 
may in turn suppress turbulence and facilitate ITB formation \cite{Gunter2001mhd,pinches2001fishbone,liuSimpleModelInternal2023}. 
Recent gyrokinetic simulations have shown that self-generated ZFs 
can dominate the nonlinear saturation of the fishbone and are linked to ITB formation in DIII-D \cite{Brochard2024,brochard2025}. 
However, direct experimental evidence of fishbone-driven ZFs has remained elusive, 
primarily due to the lack of core flow diagnostics with sufficient spatial and temporal resolution. 
Thus, whether the fishbone instability indeed drives flows that enhance confinement remains an open question. 

In this Letter, we report the first direct experimental identification of $\boldsymbol{E} \times \boldsymbol{B}$ flows 
driven by the fishbone instability in the EAST tokamak, 
using multi-channel Doppler reflectometry with high temporal and radial resolution 
\cite{gaoUpgradesWbandDoppler2025,gaoMeasurementsVelocityFluctuation2025}. 
These flows are nonlinearly excited via the self-coupling process of the fishbone, as confirmed by bispectral analysis. 
Remarkably, the flows exhibit a fine-scale and radially reversed structure within the $q=1$ surface, 
contrasting sharply with the macroscopic, well-like pattern reported in previous simulations \cite{Brochard2024,brochard2025}. 
The flow rises faster and saturates earlier than the fishbone within a single burst, 
favoring a beat-driven process over the EP expulsion model 
(Nevertheless, the possibility of additional contributions from EP expulsion at later stages is not excluded). 
Global nonlinear gyrokinetic simulations with the GTC \cite{Lin1998} quantitatively reproduce the observed radial profile 
and reveal that the fine reversed structure arises from 
the cancellation of comparable but opposite contributions from thermal ions and electrons, 
while energetic ions provide only a modest large-scale contribution. 
These findings establish the fishbone as an effective generator of sheared ZFs.

\begin{figure}[htb]
	\includegraphics[width=\linewidth]{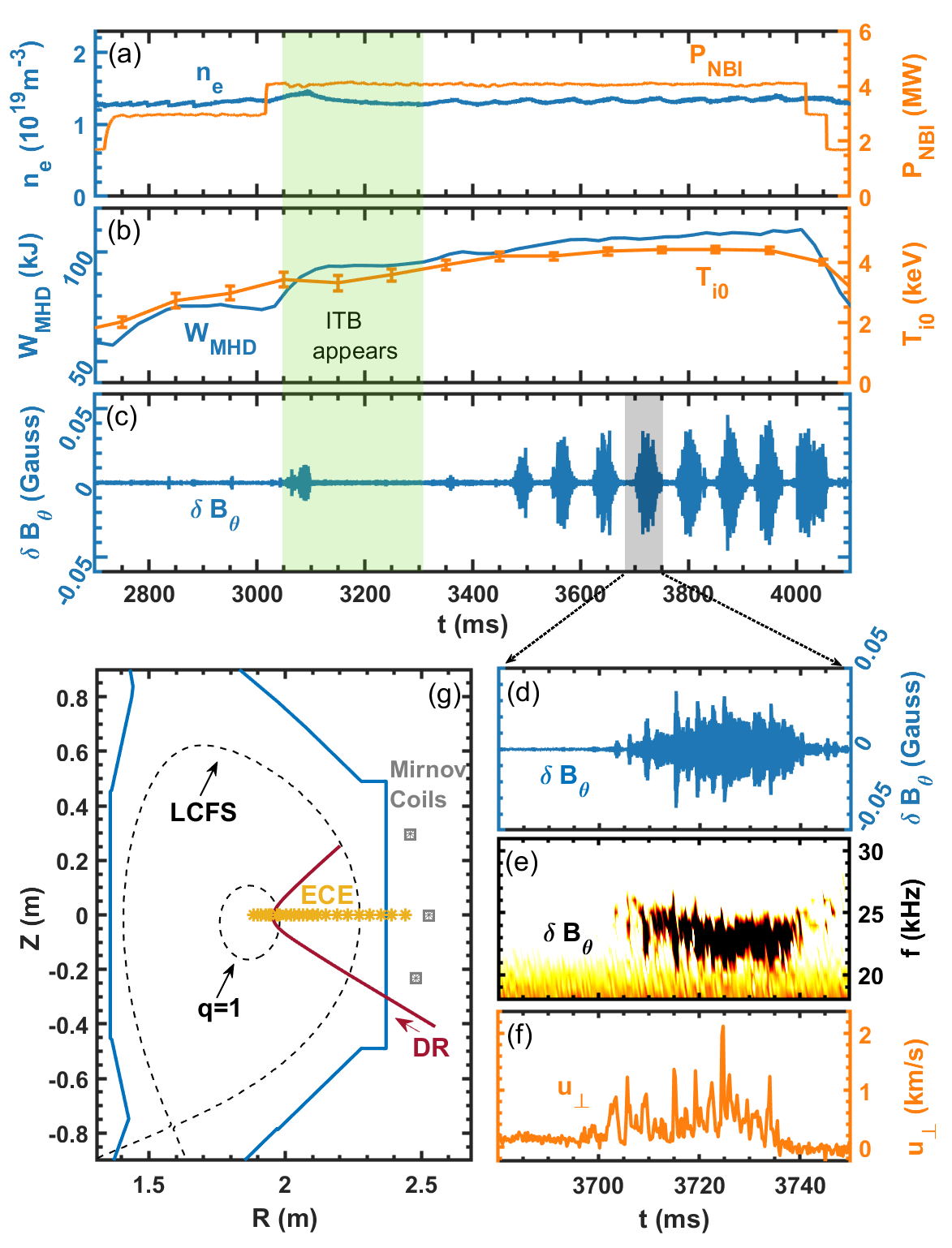}
	\caption{\label{fig1}
		Time evolutions of (a) electron density (blue) and NBI power (orange), (b) stored energy (blue) and ion temperature on axis (orange), 
		(c) edge poloidal magnetic field fluctuations $\delta B_\theta$ (filtered in the fishbone frequency range). The green shaded region denotes the establishment of the ITB. (d) presents the enlarged view of the gray shaded region in (c), and (e) displays the corresponding spectrogram, and (f) shows the corresponding low-pass filtered perpendicular velocity $u_\perp$. (g) illustrates the cross-section view of EAST and key diagnostics utilized in this Letter.
	}
\end{figure}

\paragraph{\label{sec:setup} Experimental setup.---}  
EAST is a superconducting divertor tokamak with major radius $R_0\approx1.89~\mathrm{m}$ and minor radius $a\approx0.45~\mathrm{m}$. 
The cross-section view of EAST is presented in Fig.~\ref{fig1}(g), 
where the $q=1$ rational surface and the last closed flux surface (LCFS) are indicated. 
Key diagnostics utilized in this study are also illustrated in Fig.~\ref{fig1}(f): 
Doppler reflectometry (DR) for $\boldsymbol{E} \times \boldsymbol{B}$ flows, electron cyclotron emission (ECE) for electron temperature and Mirnov coils for magnetic fluctuations. 
The experiment was conducted in low density plasmas heated solely by neutral beam injection (NBI), with toroidal magnetic field $B_0\approx \pm 2.45~\mathrm{T}$, plasma current $I_p\approx (450\sim550)~\mathrm{kA}$ and 
line-averaged electron density $n_e\approx (1\sim2)\times10^{19}~\mathrm{m}^{-3}$. 
Figs.~\ref{fig1}(a) and \ref{fig1}(b) display the time evolutions of selected experimental signals from a representative discharge \#133556. 
After full NBI power is applied at $3~\mathrm{s}$, an ITB gradually forms, enhancing core confinement. 
From approximately $3.45~\mathrm{s}$, the fishbone instability bursts periodically, 
as seen from the edge poloidal magnetic fluctuations $\delta B_\theta$ [Fig.~\ref{fig1}(c)]. 
The enlarged view of the gray shaded region in Fig.~\ref{fig1}(c) is presented in Fig.~\ref{fig1}(d), 
and its corresponding spectrogram [Fig.~\ref{fig1}(e)] clearly displays the characteristic frequency chirping of fishbone. 
Fig.~\ref{fig1}(f) plots the low-pass filtered perpendicular velocity from DR in this time period. 
The perpendicular velocity $u_\perp$, directly linked to the $\boldsymbol{E} \times \boldsymbol{B}$ flow \cite{conwayPlasmaRotationProfile2004}, 
exhibits clear coincidence with the burst of fishbone. 

\begin{figure}[htb]
	\includegraphics[width=1\linewidth]{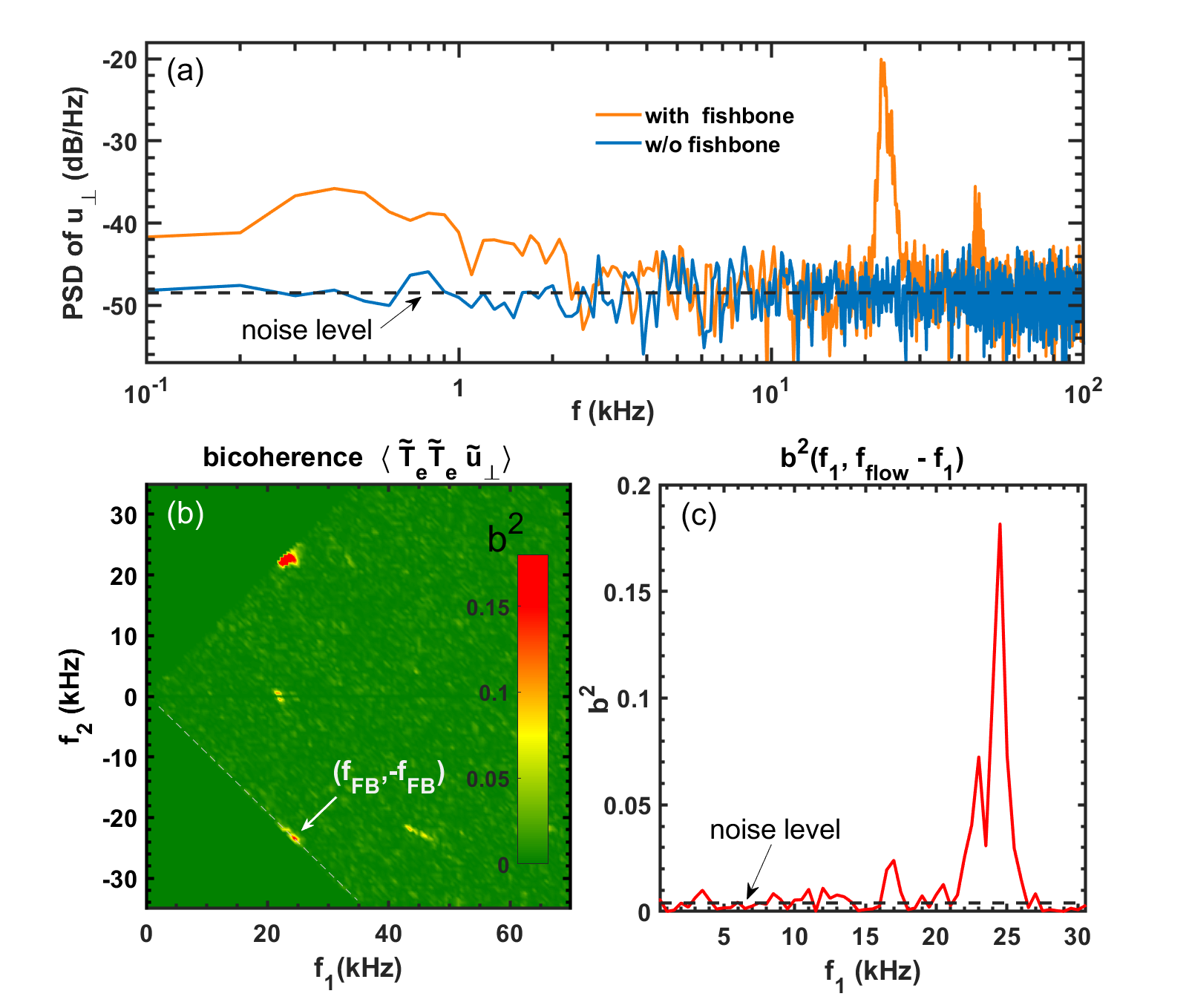}
	\caption{\label{fig2}
		(a) shows the power spectral density of $u_\perp$ with fishbone (orange) and without fishbone (blue). 
		(b) is the bicoherence $\langle \widetilde{T}_e \widetilde{T}_e \widetilde{u}_\perp \rangle$. (c) is the gray dashed line in (b). 
		The black dashed lines in (a) and (c) represent the background noise level.
	}
\end{figure}

\paragraph{\label{sec:beatdriven} Flows driven by fishbone.---}
The generation of $\boldsymbol{E} \times \boldsymbol{B}$ flows by fishbone is evidenced in Fig.~\ref{fig2}. 
A pronounced enhancement of low-frequency flows during fishbone bursts is observed, 
as demonstrated by the power spectral density (PSD) of $u_\perp$ in periods with and without fishbone [Fig.~\ref{fig2}(a)]. 
To identify the role of fishbone in exciting these low-frequency flows, 
bispectral analysis is applied [Fig.~\ref{fig2}(b) and \ref{fig2}(c)], 
a standard technique for detecting nonlinear three-wave coupling \cite{kimDigitalBispectralAnalysis1978}. 
The bicoherence $b^2(f_1,f_2)$ is defined as:
\begin{equation*}
	b^2(f_1,f_2) = \frac{\left | \langle F_1(f_1)F_2(f_2)F_3^*(f_1+f_2)\rangle\right |^2}
	{\langle\left| F_1(f_1)F_2(f_2) \right|^2 \rangle \langle\left| F_3^*(f_1+f_2) \right|^2 \rangle},
\end{equation*}
where $\langle ...\rangle$ represents the ensemble average, and $F(f)$ represents the Fourier transform. 
A bicoherence peak $b^2(f_1,f_2)$ significantly above the noise background statistically demonstrates phase coherence among the modes at $f_1$, $f_2$ and $f_1+f_2$, providing evidence for nonlinear three-wave coupling.
Fig.~\ref{fig2}(b) presents a typical cross-bispectrum
$\langle \widetilde{T}_e \widetilde{T}_e \widetilde{u}_\perp \rangle$, 
namely, that of the electron temperature and the electron temperature with the perpendicular velocity, 
computed using $250$ windows with length $T=2~\mathrm{ms}$ and $50~\%$ overlap. 
The gray dashed line in Fig.~\ref{fig2}(b), marking frequencies satisfying $f_1+f_2 = f_{flow}$, 
is extracted and shown in Fig.~\ref{fig2}(c). 
A clear peak is observed at $(f_1,f_2) = (f_{FB},-f_{FB})$, 
where $f_{FB}$ is the fishbone frequency,  
providing direct evidence that 
low-frequency flows are excited via nonlinear self-coupling of fishbone. 
Following the bispectral criterion of Kim and Powers \cite{kimDigitalBispectralAnalysis1978}, 
the observed phase matching further implies the corresponding wavenumber matching ($k_{FB}-k_{FB}=0$), 
pointing to the azimuthal symmetry of the self-generated flows.

\begin{figure}[htb]
	\includegraphics[width=1\linewidth]{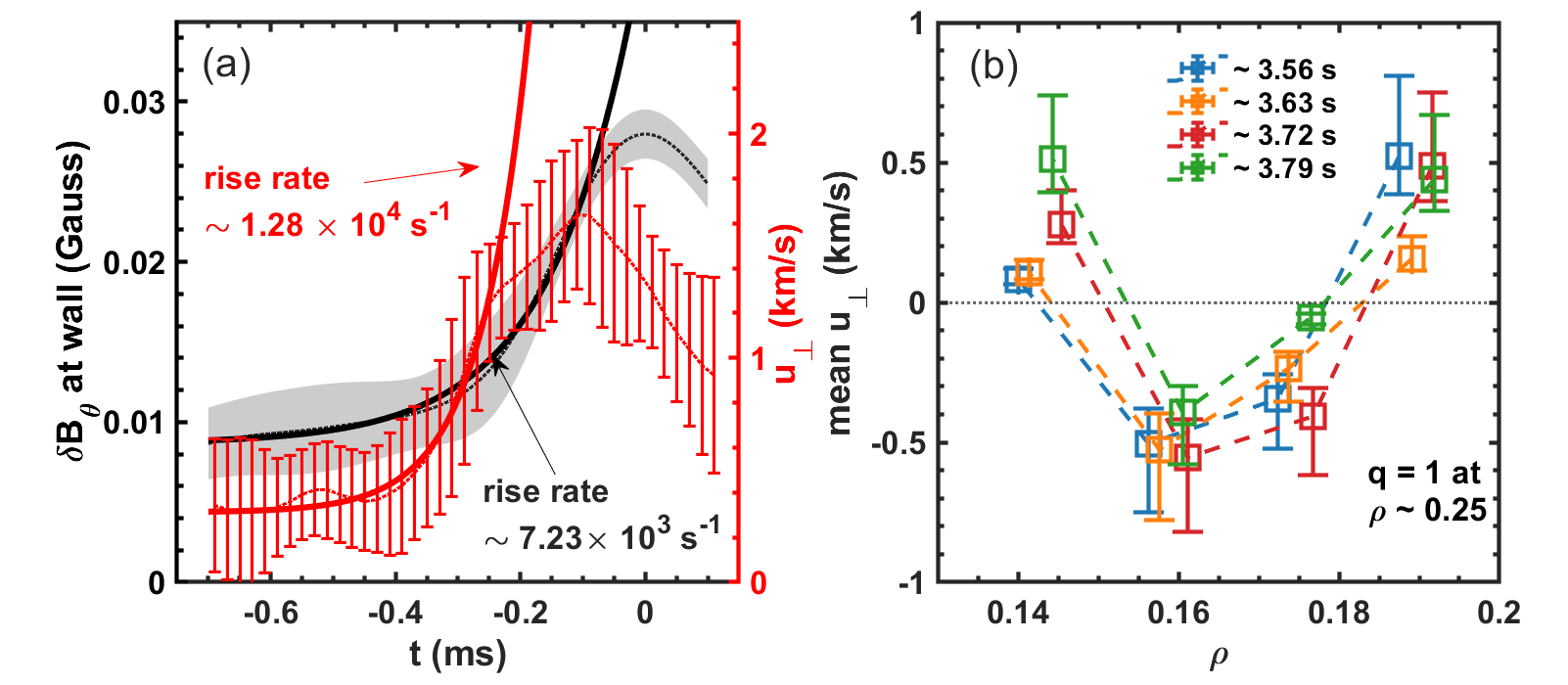}
	\caption{\label{fig3}
		(a) Ensemble-averaged temporal evolutions of low-frequency flows (red) and fishbone (black). 
		(b) Radial structure of low-frequency flows inside the $q=1$ resonant surface.
	}
\end{figure}

Temporal and radial characteristics of the fishbone-driven flows are presented in Fig.~\ref{fig3}. 
Ensemble-averaged temporal evolutions of flows and fishbone [Fig.~\ref{fig3}(a)] 
are derived from 10 fishbone bursts of comparable amplitudes, 
with the standard deviation across bursts indicating the uncertainty.
Time zero ($t=0$) corresponds to the peak of the $B_\theta$ signal. 
The rise rates of flows and fishbone are obtained by exponentially fitting $u_\perp$ and the envelope of $B_\theta$ respectively 
\cite{heidbrinkCharacterizationOffaxisFishbones2011,zhuDependenceFishboneCycle2020}.  
The temporal evolution of flows exhibits a two-stage feature. 
In the initial stage, the flow rises with a rate that exceeds the concurrent rise rate of the fishbone.
Subsequently, the flow rise rate decelerates, and the flow peaks earlier than the fishbone. 
This two-stage evolution suggests that multiple mechanisms drive the flow excitation, 
with their relative dominance shifting between stages, 
considering that the flow damping rate should remain approximately constant during a single fishbone burst under unchanged equilibrium parameters. 
Fig.~\ref{fig3}(b) shows the radial flow structure inside the $q=1$ resonant surface for the reference shot \#133556, 
measured simultaneously across multiple channels of the DR.
The mean flow velocities, obtained with an averaging window of $10~\mathrm{ms}$, 
are below $1\mathrm{km/s}$. 
What merits particular attention is that flows exhibit different directions, 
constituting a fine structure. 
This fine-scale and direction-reversed flow structure is
not anticipated by established theories and simulations 
of fishbone-driven $\boldsymbol{E} \times \boldsymbol{B}$ flows, 
which predicted macroscopic flow structures without direction reversal 
based on the EP-expulsion mechanism 
\cite{pinches2001fishbone,liuSimpleModelInternal2023,Brochard2024,brochard2025}. 
These observations collectively indicate that the low-frequency flows are ZF-like in
nature, motivating a quantitative comparison with first-principles simulations. 

\paragraph{\label{sec:gtc}Zonal electric field driven by Fishbone.---}
Gyrokinetic simulations of EAST discharge \#133556 are carried out with GTC. 
The magnetic equilibrium at $t=3.72$ s is reconstructed using kinetic-EFIT,
and plasma profiles are obtained from TRANSP simulations constrained by experimental measurements.
In this discharge, the total plasma $\beta_t$ is $1.9\%$,  
where $\beta$ denotes the ratio of plasma pressure to magnetic pressure.
The energetic-ion density profile is obtained from NUBEAM, 
and we use the slowing-down distribution in the simulation
with a neutral beam injection (NBI) energy of approximately $45\,\mathrm{keV}$.
On the magnetic axis, the energetic-ion beta fraction and density ratio are 
$\beta_f/\beta_t = 0.21$ and $n_f/n_e = 0.13$, respectively, where $n_e$ is the electron density.
All nonlinear simulations cover the core region which includes the magnetic axis, 
and extend to $\rho = \sqrt{\psi_T/\psi_{T,\text{edge}}} = 0.78$ in GTC,
where $\psi_T$ denotes the toroidal flux.
Physical constraints and Fourier series are 
employed on the perturbation \cite{Lewis1990physical},
resolving singularity issues associated with the magnetic axis.
Beyond this radial domain, an outer edge buffer is employed in which the equilibrium gradients are removed.
We employ a $\delta f$ method for gyrokinetic thermal and energetic ions, 
and treat electrons with a fluid-kinetic model.
Numerical convergence has been carefully verified in the GTC simulations. 
The simulation parameters include radial grids with $N_{\psi} = 100$,
poloidal grid points $N_{\theta} = 400$, and a field-aligned mesh consisting of $N_{\|} = 32$ parallel grids.
Nonlinear electromagnetic runs employ 400 marker particles per cell for each of the three species.

\begin{figure}[htb]
	\centering	
	\begin{overpic}
		[scale=0.41]{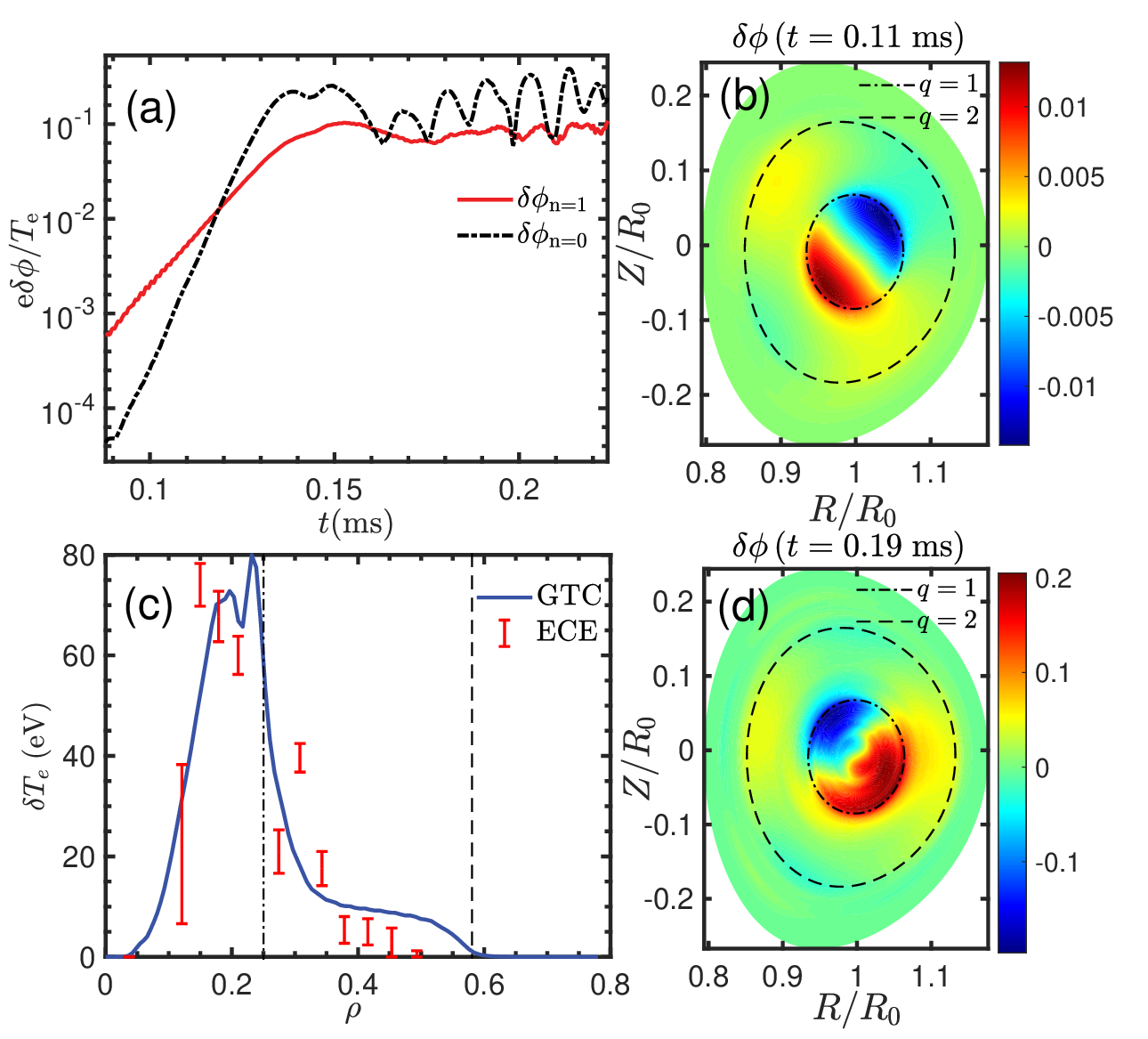}
	\end{overpic}	
	\caption{
		(a) Time evolution of the perturbed electrostatic potential $\delta\phi$ 
		(normalized by electron temperature $T_e$ and charge $e$) from fishbone nonlinear gyrokinetic simulations.
		The black dashed lines represent the zonal potential $\phi_{00}$, 
		shown as root-mean-square (rms) values across the radial simulation domain.	
		Poloidal contour plots of $\delta\phi$ for the fishbone
		simulations during the linear (b) and nonlinear (d) phases, respectively. 
		(c) $\delta T_e$ at $t = 0.19\,\mathrm{ms}$ for fishbone simulations, 
		compared with ECE diagnostic results, 
		and the vertical lines marking the $q=1$ and $q=2$ rational surfaces, respectively. 
	}
	\label{fig4}
\end{figure}

To investigate the nonlinear saturation of the EP-driven fishbone observed in this EAST discharge, 
we carry out global nonlinear GTC simulations that retain only the $n=1$ toroidal mode together with the $n=m=0$ ZFs.
Figure \ref{fig4}(a) shows the time evolution of 
the perturbed electrostatic potential $\delta\phi$ and zonal potential $\delta\phi_{00}$.
The $n=1$ fishbone grows with a linear growth rate $\gamma_{n=1} = 9.78\times 10^4~\mathrm{rad/s}$ 
and mode frequency $f = 20.8~\mathrm{kHz}$.
During the linear and intermediate phases, $\delta\phi_{00}$ 
increases exponentially at a rate $\gamma_{n,m=0} \approx 2\,\gamma_{n=1}$ 
, indicating that the ZFs are generated by fishbone through the beat-driven process \cite{Chen2024}.
The behavior that 
the ZFs saturate earlier than the $n=1$ fishbone mode is consistent with experimental observations [Fig.~\ref{fig3}(a)]. 
The fishbone instability saturates at $t \approx 0.19\,\mathrm{ms}$, 
with a normalized magnetic perturbation amplitude of $\delta B_{\perp}/B_0 \sim 3 \times 10^{-3}$ 
in the presence of ZFs.
The spatial structure of the $n=1$ electrostatic potential is shown in Figs. \ref{fig4}(b) and \ref{fig4}(d). 
In the linear phase, the mode exhibits a typical $n=m=1$ structure 
localized predominantly inside the $q=1$ rational surface. 
Entering the nonlinear stage, the poloidal structure becomes twisted and radially broadened 
due to nonlinear coupling with the EP drive and ZFs.
As shown in Fig. \ref{fig4}(c), a direct comparison between GTC simulations 
and electron cyclotron emission (ECE) measurements validates the saturation level of the fishbone instability.
The electron temperature perturbation envelope $\delta T_e$ from the nonlinear GTC simulation at saturation 
is in good agreement with the ECE measurements.

\begin{figure}[htb]
	\centering	
	\begin{overpic}
		[scale=0.36]{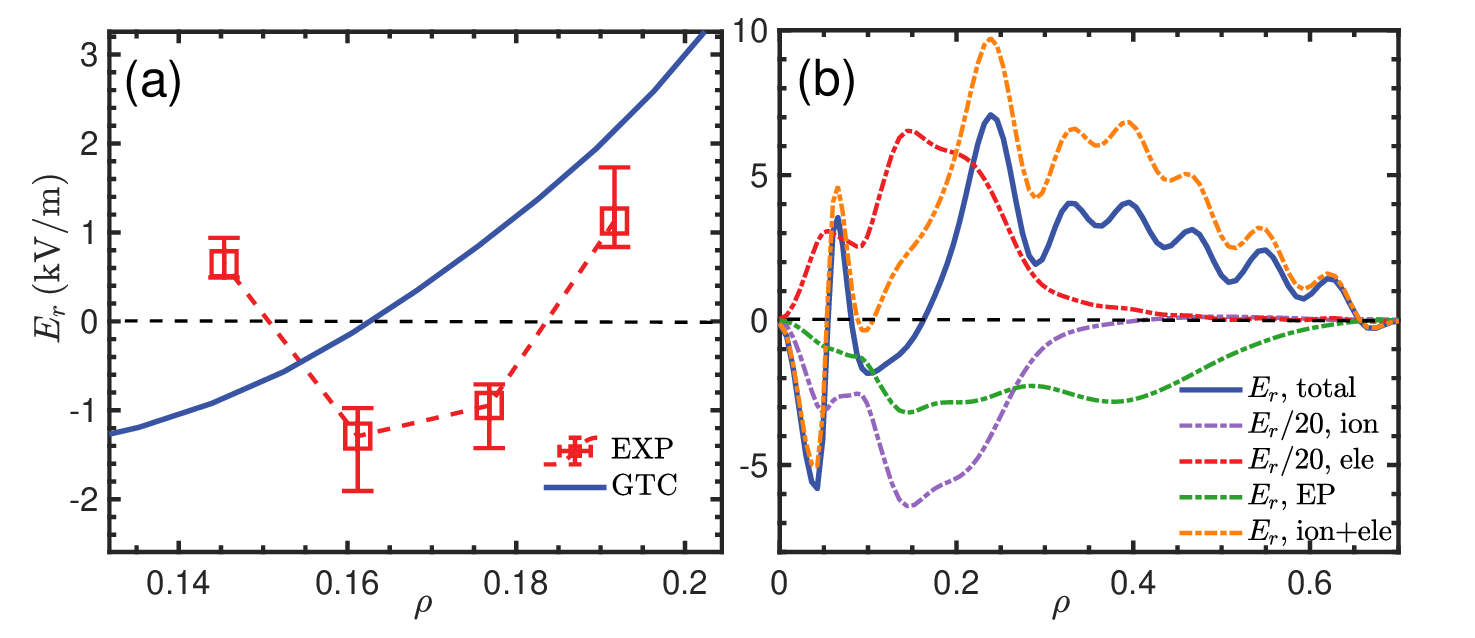}
	\end{overpic}	
	\caption{
		(a) Radial electric field $E_r$ driven by the fishbone in GTC simulation and 
		DR measurements.
		(b) Contributions of thermal ions, thermal electrons and energetic ions to the simulated fishbone-driven $E_r$, 
		contrasting their roles in the field structure.	
	}
	\label{fig5}
\end{figure}

The observed $\boldsymbol{E} \times \boldsymbol{B}$ flow originates 
from a zonal radial electric field nonlinearly generated by the fishbone, as shown in Fig.~\ref{fig5}(a). 
The measured radial profile from multi-channel DR agrees quantitatively with the GTC simulation result, 
where the simulation is performed for a representative burst event.
The field exhibits a radially reversed structure within the $q=1$ rational surface, 
with a radial scale close to the microscopic and mesoscopic scales characteristic of ZFs 
driven by drift-wave turbulence and Alfv\'en eigenmodes \cite{chen2016physics,diamondZonalFlowsPlasma2005}. 
This differs from the macroscopic, well-like zonal electric field reported in previous simulations \cite{Brochard2024}. 
The difference is attributed to the self-consistent inclusion of the zonal electron density response in our simulation model.
It also contrasts with the equilibrium radial electric field derived from radial force balance, 
which exhibits a broad, macroscopic scale profile dominated by the ion temperature gradient.

To elucidate the origin of fine and reversed structure, 
we separate the contributions from thermal ions, thermal electrons, 
and energetic ions by solving the flux surface averaged gyrokinetic Poisson equation 
for each species to obtain the corresponding contribution to $E_r$, as shown in Fig.~\ref{fig5}(b).
The contributions from thermal ions and electrons to the $E_r$ 
are comparable in magnitude but opposite in direction.
The zonal electron density response is therefore crucial, 
as it partially cancels the zonal ion density, 
reducing the net charge separation and thus significantly weakening the total amplitude of $E_r$. 
The radially reversed structure is determined primarily by the interplay of thermal ions and electrons, 
not by energetic ions.
Although the individual field contributions from either thermal species alone are much larger than that from energetic ions, 
their mutual cancellation results in a relatively small net field.
In contrast, the energetic ions provide only a modest contribution, 
predominantly in the long-wavelength ($k_r \simeq 0$) component, 
and do not determine the observed mesoscopic radial structure.

\paragraph{\label{sec:discussion} Discussion.---}

In summary, we have presented the first direct experimental identification of $\boldsymbol{E} \times \boldsymbol{B}$ flows 
driven by the fishbone instability in the EAST tokamak core, 
enabled by high-resolution multi-channel Doppler reflectometry. 
The observed flow exhibits a fine-scale, radially reversed structure within the q=1 surface. 
Global nonlinear gyrokinetic simulations with GTC quantitatively reproduce the measured radial profile 
and confirm that the flow originates from a zonal radial electric field nonlinearly generated by the fishbone. 
The simulated $\delta T_e$ envelope also agrees well with ECE measurements, validating the saturation level of the instability. 
A species-decomposition analysis of the gyrokinetic Poisson equation reveals that 
the reversed fine structure is primarily determined by the comparable but opposite contributions of thermal ions and thermal electrons, 
while energetic ions provide only a modest, long-wavelength contribution. 
This result is not captured by the energetic-ion-expulsion picture 
\cite{Gunter2001mhd,pinches2001fishbone,liuSimpleModelInternal2023,Brochard2024,brochard2025}, 
which treats electrons as a passive background and predicts a macroscopic, 
energetic-ion-dominated flow without direction reversal, 
thereby highlighting the essential role of thermal-species kinetics 
in the observed fine-scale reversal. 

The temporal evolution within a single burst, 
where the flow rises faster and saturates earlier than the fishbone, provides additional insight into the excitation mechanism. 
This behavior is more consistent with the beat-driven process than with the EP expulsion picture, 
in which the flow growth would be expected to track the instability more closely \cite{liuSimpleModelInternal2023}. 
Importantly, these two mechanisms are not mutually exclusive. 
As shown in Fig.~\ref{fig5}, EPs do generate a macroscopic contribution to $E_r$, 
which may become more prominent at later stages or under stronger EP redistribution. 
Our results thus suggest a two-stage picture: beat-driven excitation dominates the early flow generation, 
while EP-expulsion-driven flows may emerge on longer time scales. 
This perspective reconciles our observations with previous reports linking fishbone bursts to ITB formation 
in ASDEX Upgrade, HL-2A and EAST \cite{Gunter2001mhd,Chen2016dynamics,Deng2022investigation,Gao2018sustained,Liu2020experimental,Zhang2022}, 
where both mechanisms could contribute synergistically.

The fine-scale, radially sheared flows identified here 
offer a promising pathway to regulate core turbulence while largely preserving EP confinement, 
since beat-driven flows do not intrinsically rely on EP loss. 
Future work will systematically investigate the role of these fishbone-driven flows in ITB formation and sustainment, 
as well as their interplay with background turbulence across multiple spatial scales.

The authors acknowledge Z.Y. Qiu and Z.Y. Liu for their useful discussions.
This work is supported by the National MCF Energy R\&D Program of China (Grant No. 2024YFE03050002), 
National Natural Science Foundation of China (Grant Nos. 12375230, 12475228 and 12525511), 
the Strategic Priority Research Program of Chinese Academy of Sciences (Grant Nos. XDB0500302 and XDB0790202),
and the Youth Innovation Promotion Association CAS No. 2023470.
The numerical calculations in this paper were performed on the Hefei Advanced Computing Center.

\bibliographystyle{apsrev4-2}  

\bibliography{bibliography.bib}

\end{document}